\documentclass[preprint,12pt]{aastex}

\newcommand{\degree}{\mbox{$^{\circ}$}}

\newcommand{\lsun}{\mbox{L$_\odot$}}

\input{epsf}

\def\plotfiddle#1#2#3#4#5#6#7{\centering \leavevmode
\vbox to#2{\rule{0pt}{#2}}
\includegraphics{#1}}

\begin{document}

\slugcomment{Accepted Astrophysical Journal, Jul 19, 2010}

\title {\bf A Spitzer Search For Planetary-Mass Brown Dwarfs With Circumstellar Disks: Candidate Selection}
\author{Paul M. Harvey\altaffilmark{1},
Daniel T. Jaffe\altaffilmark{1}
Katelyn Allers\altaffilmark{2},
Michael Liu\altaffilmark{3}
}

\altaffiltext{1}{Astronomy Department, University of Texas at Austin, 1 University Station C1400, Austin, TX 78712-0259;  pmh@astro.as.utexas.edu, dtj@astro.as.utexas.edu}
\altaffiltext{2}{Department of Physics and Astronomy, Bucknell University, Lewisburg, PA 17837;  k.allers@bucknell.edu}
\altaffiltext{3}{Institute for Astronomy, University of Hawaii, 2680 Woodlawn Dr., Honolulu, HI 96822; mliu@ifa.hawaii.edu}

\begin{abstract}

We report on initial results from a {\it Spitzer} program to search for very low-mass brown dwarfs in
Ophiuchus.  This program is an extension of an earlier study by Allers et al. which had resulted in
an extraordinary success rate, 18 confirmed out of 19 candidates. Their program combined near-infrared and
Spitzer photometry to identify objects with very cool photospheres together with circumstellar
disk emission to indicate youth.  Our new program has obtained deep IRAC photometry of a 0.5 deg$^2$
field that was part of the original Allers et al. study.  We report 18 new candidates whose
luminosities extend down to 10$^{-4}$ \lsun\ which 
suggests masses down to $\sim$ 2 $M_J$ if confirmed.  We describe our selection techniques,
likely contamination issues, and follow-on photometry and spectroscopy that are in progress.

\end{abstract}


\section{Introduction}\label{intro}

Young brown dwarfs exhibit ``circumstellar'' disk phenomena much like their more massive
counterparts, e.g. \citep{apai04}, \citep{luhman10}.  Although there are quantitative differences in detailed physical parameters
for disks around sub-stellar objects \citep{pascucci09}, the distribution of disk properties 
appears to be relatively continuous across the sub-stellar boundary, e.g. \citet{scholz09}.  The lower limit
to the sub-stellar mass of objects that form with accretion disks is, however, still uncertain.
Theoretical considerations of the opacity limit for fragmentation of clouds, e.g. \citet{bate09}, suggest that such
a formation mechanism is likely to be limited to central objects above a few Jupiter masses (M$_J$).
In recent years a number
of studies have been made to push the detection limit on low-mass brown dwarfs
to the lowest possible levels, in part to test these theoretical predictions.
In general, these studies have taken advantage of the fact that young BD's are significantly
more luminous and detectable than older field BD's, and thus these investigations have
focussed on star-forming regions.  Infrared excesses due to circum-object material provide an
additional discriminant to isolate young brown dwarfs, as well as a tool to investigate the properties of
disks around such cool, low-mass central objects.


One of the most successful searches for young, low-mass BD's was that of \citet{allers06}, hereafter A06.  Since the
original publication of their 19 candidates, \citet{allers07} and \citet{gully10}
have spectroscopically confirmed 18 out of the 19 candidates as low-mass stars and sub-stellar objects.
A06 used a combination of color and magnitude criteria
to select objects with the properties of cool, low-luminosity sub-stellar objects in fields
observed by the Spitzer {\it c2d} Legacy program \citep{evans03}.  The Spitzer data provided the ability
to search for infrared excesses, presumably due to circumstellar disks, which would imply
youth and membership in the star-forming clouds.
The observational limit on low-luminosity objects in their study was set by the {\it c2d}
sensitivity in the two longest wavelength Spitzer IRAC bands at 5.8 and 8.0\micron, a result
of the relatively short integration times of 48 sec required by the large area survey.

We report here on a program of much deeper Spitzer imaging of a 0.5 deg$^2$ area in
the Ophiuchus star-forming region where matching, deep I, J, H, and K$_s$ photometry
already exist from the original A06 study.  Our program was aimed at
detecting young, planetary-mass objects down to masses of 2 M$_J$ with circumstellar disks.
The youngest YSOs in the core of Ophiuchus are probably under 1 Myr in age, but star formation appears to have
been underway for 2 - 5 Myr in the more extended area surveyed in our study \citep{wilking08}.
In addition to testing the opacity limit for fragmentation and the lower limit of the IMF, detection of a number of such
objects would provide a sample to: 1) study the properties of disks physics in an extended range of
temperature and gravity parameter space; 2) probe the atmospheres of the lowest mass 1 Myr old BD's; 3) test the
\.{M}/M$_*$ relation to lower masses; and 4) examine the distribution and multiplicity of very low mass objects for
hints to their origin.
In this first paper, we focus on objects that were pre-selected from the IJHK data to 
meet all the selection criteria of A06 for the faintest magnitude bin.  We then
examined our observations of these objects to search for examples with IRAC excesses.

In \S\ref{obs} we describe the details of our Spitzer observations and data reduction processes.
We then discuss how we have selected our candidate objects  in \S\ref{select}.   In \S\ref{anal}
we discuss the kinds and number of contaminants that we expect in our sample
and finally mention followup observations that are planned to investigate the reliability of
this sample.

\section{Observations and Data Reduction}\label{obs}


Table \ref{aortab} lists the AOR's that were part of this Spitzer program PID 50025.  
The total integration time per pixel was roughly
900 sec in all four IRAC bands.  The three shorter wavelength bands were observed with
100 sec frame times, while band 4 (8.0\micron) was observed with 50 sec frame times.


We used most of the features of the {\it c2d} pipeline \citep{evans07} to process the data.  We
did not, however, use any additional masking beyond the recommended mask bits in the
IRAC data handbook\footnote{http://ssc.spitzer.caltech.edu/irac/iracinstrumenthandbook/home/}, and band-merging was accomplished with a simple positional-matching
algorithm, i.e. positional coincidence within 2 arcsec. Sources with any significant confusion were hand-checked
to be sure the matches were reliable.
The frames were mosaicked with the standard version of the Spitzer mosaicker, Mopex, with
the output pixel size set to be equal to the input pixel size.  The {\it c2d} source 
extraction tool, c2dphot \citep{evans07}, was used in several different modes to extract sources and 
measure fluxes.  For most of the objects we used the standard mode of extraction from the
full mosaic at each wavelength, and then flux measurement from the stack of individual BCD's
appropriate for each source.  For a few of the final selected sources, the level of diffuse
background was so high, that even with the PSF-fitting used by c2dphot, we were unable to
extract a source at 5.8 or 8\micron.  In these cases 
we used the tool in a mode where the position was held fixed at the source position determined from
the IRAC band 1 and 2 extractions while the flux was determined from a PSF-fit.  Tests we
have done show that this can underestimate the true flux by order up to 20\% because the
position is not allowed to vary during the fitting.  In any case, the fluxes of all these fixed-position
sources have relatively high uncertainties due to the high background.  Since these two
longest wavelengths were used to search for circumstellar excess in our program, the
underestimate of the fluxes means that our final list of excesses is likely to be
slightly conservative if anything.

\section{Candidate Selection}\label{select}

The goal of this first part of our study was to look specifically at objects in Ophiuchus that
had been selected by the criteria in A06 for the faintest magnitude bins and which
were lacking IRAC detections at 5.8 and 8.0\micron\ due to the modest integration
times of the {\it c2d} survey.   Using those criteria resulted in a list of 605
objects in the area covered by A06 at I--K$_S$.  Of these, 549 were in the area
covered by our new Spitzer observations at both 5.8 and 8.0\micron\, and 578 were
covered at 5.8\micron.
Of these, 179 sources were bright enough for successful flux extraction.
The existing A06 shorter wavelength data were then band-merged with our new photometry
for these sources.  In a future study we will take a broader look at our photometry beyond
this first set of candidates.

\begin{figure}
\plotfiddle{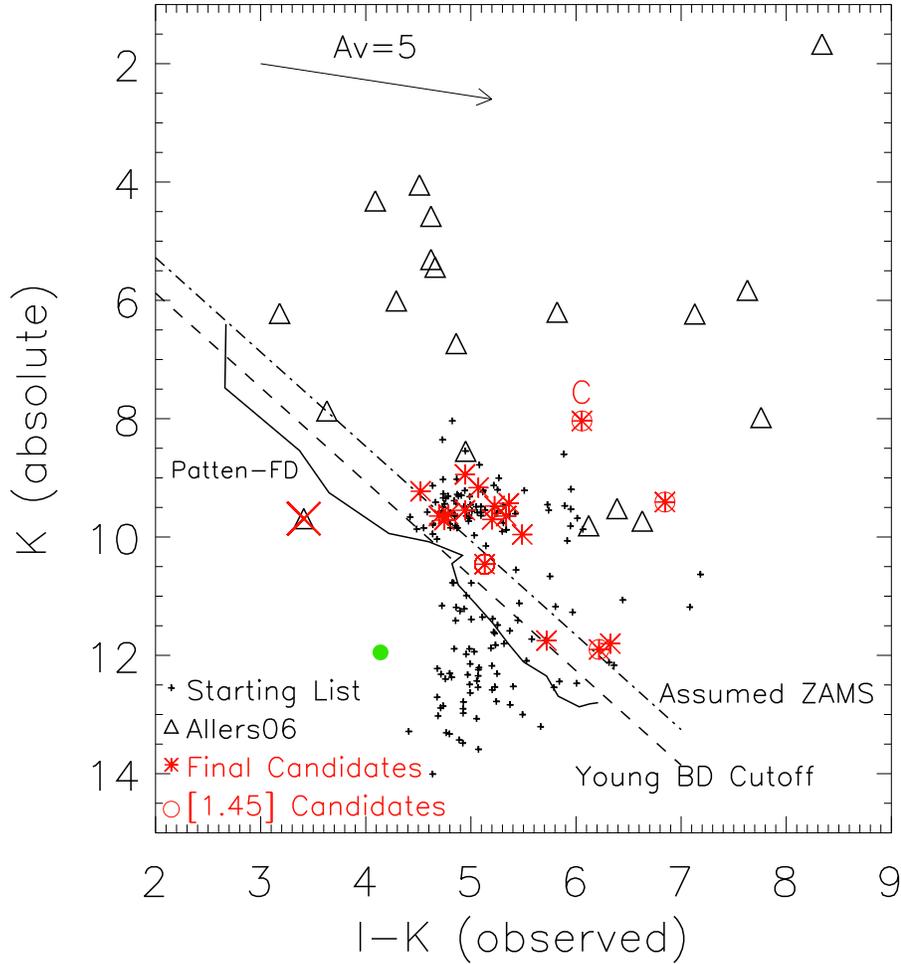}{5.0in}{0}{80}{80}{-300}{-200}
\figcaption{\label{ikkf}
Absolute K$_S$ magnitude versus I -- K$_S$ color (assuming a distance of 125pc for all objects).
Our final selected candidates are marked with red asterisks; the 161 candidates out of our starting
list of 179
candidates that did not meet all the selection criteria are marked by small crosses; objects that were found by Allers
and Liu (2010) with likely water absorption in the near-infrared are marked with red circles; the one
source in our candidate list that has already been spectroscopically confirmed is marked also with a ``C'';
and the locations in this diagram of the original A06 candidates are marked with black triangles.
The one source from A06 that was found to be extragalactic is overlaid with a large red ``X''.
A solid green circle shows the approximate location in this color-magnitude space of the 2 M$_J$ diskless model
from A06.
The solid line shows the distribution of field dwarfs in this color-magnitude space from \citet{patten06} and
\citet{cabal06}.  The
dash-dot line is a straight line about 1 mag brighter than the Patten distribution, which we chose as a probable
distribution for unreddened young brown dwarfs in order to estimate the interstellar reddening to our candidates.
The dashed line below that marks our limit in this color-magnitude space for selecting objects in our final candidate
list that also fit our other selection criteria.}
\end{figure}

\begin{figure}
\plotfiddle{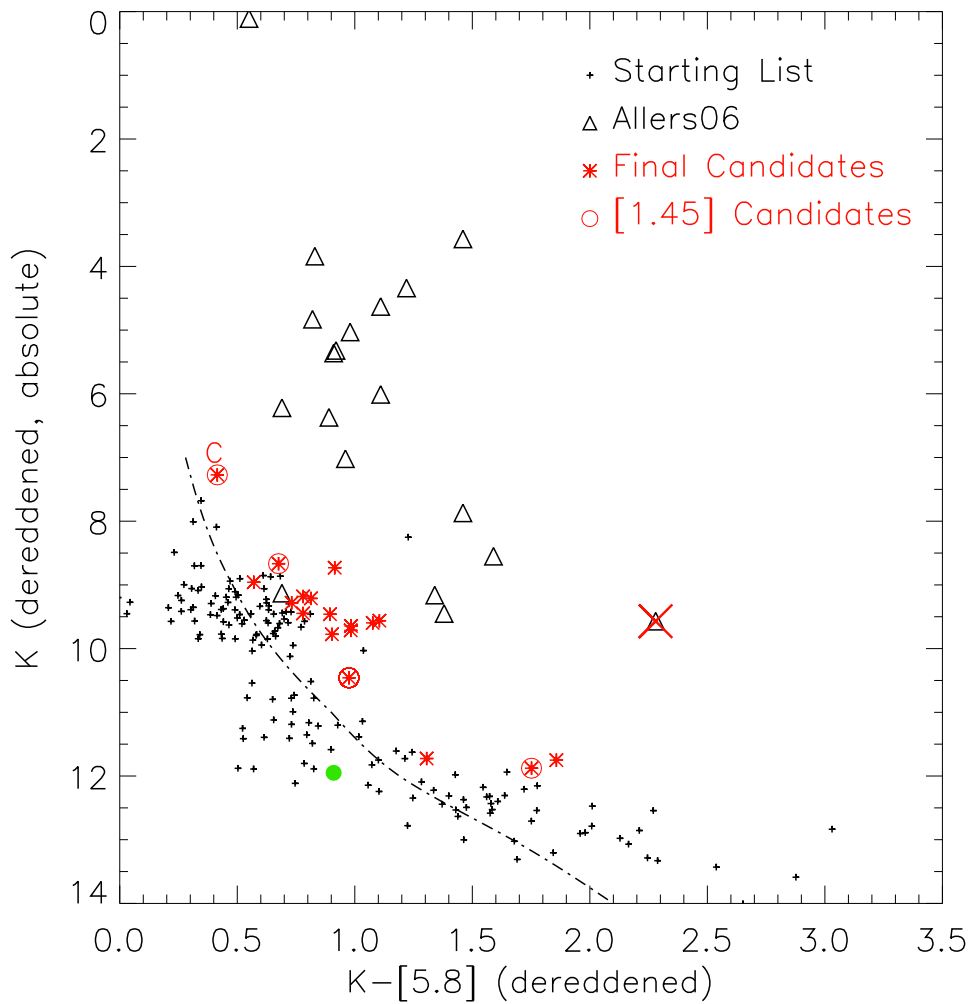}{7.0in}{0}{80}{80}{-300}{-150}
\figcaption{\label{k58}
Absolute dereddened K$_S$ magnitude versus dereddened K$_S$ -- [5.8] color as described in the text.  The symbols
are the same as in Fig. \ref{ikkf}. The dash-dot line shows the rough distribution of field dwarfs in this color-magnitude
space from \citet{patten06}.  This diagram was used to select sources with possible disk emission by choosing sources at
least 0.1 mag above the dash-dot line that also met our other criteria.}
 
\end{figure}

In order to search for candidate objects with an infrared excess, it is of course
necessary to know the colors/magnitudes of comparable objects without an excess.  This
may be problematic, however, since we cannot know for sure what the atmospheres of
such low-mass brown dwarfs look like.  Since the A06 study, \citet{patten06} have
published an extensive set of Spitzer photometry of field brown dwarfs as late
as spectral type T8.  Roughly speaking, we would expect that a young 
few M$_J$ object would have a spectral type of mid- to late-L \citep{burrows97}, but would likely be more luminous
than older field dwarfs of the same spectral type.  
Objects near the deuterium-burning limit, M $\sim$ 15 M$_J$, for comparison would have luminosities $\sim 10^{-3}$
\lsun, and spectral types of late M to early L  for roughly the first 10 Myr of their lifetimes \citep{burrows97, chabrier00}.
(Of course, recent theoretical investigations have
shown that the accretion history of young, low-mass objects can make the correspondence between mass
and observable quantities highly problematic \citep{baraffe09}).  We have, therefore, developed
two additional selection criteria beyond those of A06, to select objects with
shorter wavelength colors appropriate for planetary-mass BD's and longer wavelength
colors implying an excess above that expected for known field dwarfs.  These selection
criteria are illustrated in Figures \ref{ikkf} and \ref{k58}.

In Figure \ref{ikkf}, we have plotted the absolute K$_S$ magnitude versus the observed
I--K$_S$ color for a variety of objects.  We chose this color-magnitude pair because
at I and K$_S$ there should be little excess emission from circumstellar disks, and
the colors of field dwarfs listed by \citet{patten06} and \citet{cabal06} are nearly monotonic with spectral
type in these bands.  The absolute magnitudes of observed objects in Ophiuchus have
been calculated assuming a distance modulus of 5.48 mag.  We have used this diagram for
two purposes: 1) to provide an additional color-magnitude selection of objects with
colors of cool sub-stellar objects, but with magnitudes brighter than the older field
dwarfs of \citet{patten06}, and 2) to estimate conservative values for the reddening
to our candidates in order to more accurately select objects with longer wavelength
excesses.  The solid, somewhat wavy black line shows the colors and magnitudes for the
field dwarfs of Patten et al. from M3 to L7.  The large open triangles show the locations in this
diagram for the 19 candidates of A06 (the one that was, in fact, extragalactic is noted
by a large overlying ``X'').  The green circle shows the approximate location for the
2 M$_J$ disk-less model described in A06.

We used the combination of the Patten et al. data and the Allers et al. candidates
to set two, somewhat arbitrary, limits on the figure.
The upper dot-dash line shows our assumed color-magnitude relation for diskless, very
young low-mass brown dwarfs in order to estimate reddening to our objects.  We chose this
line simply by the fact that it is roughly a magnitude brighter at all colors than the
field dwarf relationship, and is at the bottom of the distribution of colors/magnitudes
of the confirmed objects from A06.  We believe this to be a conservative estimate of
intrinsic colors in the sense that it probably provides an upper limit to the
actual interstellar extinction; indeed, spectroscopic confirmation of the brightest of our 
candidates \citep{allers10} finds a reddening somewhat less than that estimated
from this simple relation.   
The extinction values listed in Table \ref{cand_table}
were derived by dereddening our observed colors back to this line.
The lower, dashed, line shows our cutoff for selection of
candidates in this color-magnitude space.  
This line was set, again somewhat arbitrarily,
at about a half magnitude brighter than the field dwarf relation in order to select for
young, i.e. larger diameter objects.  In addition, as described in the figure, we show
the positions of the final selected candidates as well as objects that were rejected,
either because of their position in this diagram or Figure \ref{k58}, or because
they appeared to be extended in one of the images.

Figure \ref{k58} shows our selection criterion aimed at choosing objects with some
excess emission in the longer IRAC bands to pick out objects with circumstellar disks,
which would then be more likely to be young objects in the Ophiuchus star-forming cloud.
We decided to use the 5.8\micron\ photometry as a measure of disk excess rather than
the 8.0\micron\ values because a significant number of our candidates were surrounded
by enough diffuse (presumably PAH) emission that extracting good photometry at 8.0\micron\
was highly problematic.
The symbols have the same meaning as in the preceding figure (Figure \ref{ikkf}).
The colors and magnitudes have been dereddened using the extinction derived from
Figure \ref{ikkf} which is also listed in Table \ref{cand_table}.  We assumed the same
extinction dependence on wavelength as A06 for consistency.  In this diagram we
selected any object that was more than 0.1 magnitude redder than the relation shown
for the field dwarfs of Patten et al.  Note, in this color-magnitude space the extragalactic
interloper in the A06 study is significantly redder than any of the low-mass brown
dwarfs, while in Figure \ref{ikkf} it is bluer, and in fact would have been eliminated
by these new selection criteria.

After applying both of these color-magnitude criteria to the 179 objects that were
detected at 5.8\micron\ out of the initial 605 candidates and eliminating any objects that
appeared extended in either the I or 3.6\micron\ images, we were left with 18 new
final candidates;  these are listed in Table \ref{cand_table}.
Their luminosities were derived by integrating the dereddened spectral energy
distributions from I through 8\micron\ assuming the extinctions listed.
In the following section we discuss the probability that this list actually contains
examples of planetary-mass young brown dwarfs.

Although our study is essentially an extension of that by A06, there {\it are} a few small
differences in our selection criteria.  Since we started with a list of sources that fit
the A06 selection criteria in the I thru 3.6\micron\ bands, that part of the selection is
identical.  For selection of sources with shorter wavelength colors of sub-stellar objects,
however, our criteria were more stringent with the addition of the color-magnitude cuts
shown in Figure \ref{ikkf}.  In principle this should result in a lower likelihood of
contamination by extragalactic objects; we discuss this issue in more detail in the following
section.  The second difference is that A06 insisted on a 3-sigma excess at both 5.8 and 8\micron\
over the colors of field M dwarfs.  Our criterion shown in Figure \ref{k58} is somewhat looser in
that: 1) we have only used the 5.8\micron\ excess, and 2) for the 2 sources closest to the cutoff
line the excess only amounts to roughly 1$\sigma$.

\section{Discussion}\label{anal}
\subsection{Contamination}

There are a variety of objects other than very low mass young BD's that may contaminate our sample
by mimicing the color/magnitude criteria that we have used in our selection..
These include: foreground or background dwarfs, extragalactic objects, particularly AGN, and distant background
evolved stars.  Here we examine the likelihood of contamination by each of these.  

We have
specifically chosen color and color-magnitude selection criteria that have worked very well to
pick out cool dwarfs in the A06 study.  One consequence of this selection scheme is that the most likely contaminants
may be foreground and background cool stars.  It is relatively easy to eliminate background giants as
likely contaminants because of the galactic latitude of our survey field, 18\degree.  Any red giant with
apparent magnitudes comparable to our selection criteria would be at a distance of at least 20 kpc, and
7 kpc above the Galactic Plane.
Low-mass field stars and BD's in the foreground and near-background will, of
course, have colors and magnitudes much closer to those of our selected objects with the exception of
the longer IRAC bands.  \citet{cabal06} have published a procedure for estimating 
exactly this sort of contamination in searches such as ours.  The situation for Ophiuchus is somewhat
more favorable than for the other two cases examined by \citet{cabal06}, Orion and the Pleiades, because
both the Sun and the cluster are on the same side of the Galactic plane so we never look through the
plane.  Using the prescription given by \citet{cabal06} it is clear that foreground contamination is
not an issue, since such objects would necessarily be lower luminosity BD's than those in the star-forming
cloud, and such objects are not numerous.  In our survey area, we would expect of order one such foreground
field dwarf contaminant.
Background BD's and M dwarfs are a more serious problem, since a background M5 dwarf with a few magnitudes
of visual extinction would have colors and magnitudes comparable to our I/K selection criteria for a distance
of a couple hundred pc.  For example, using the algorithm of \citet{cabal06} we estimate that of
order 20 M dwarfs could occupy the area in Figure \ref{ikkf} defined by $9 < K_{abs} < 10, 4.5 < I-K < 5.5$.  This
area contains a number of our candidates as well as several tens of objects that we have discarded. 
The one criterion that background field dwarfs would not satisfy, of course, is that for an excess in the K-[5.8]
colors shown in Figure \ref{k58}.

\begin{figure}
\plotfiddle{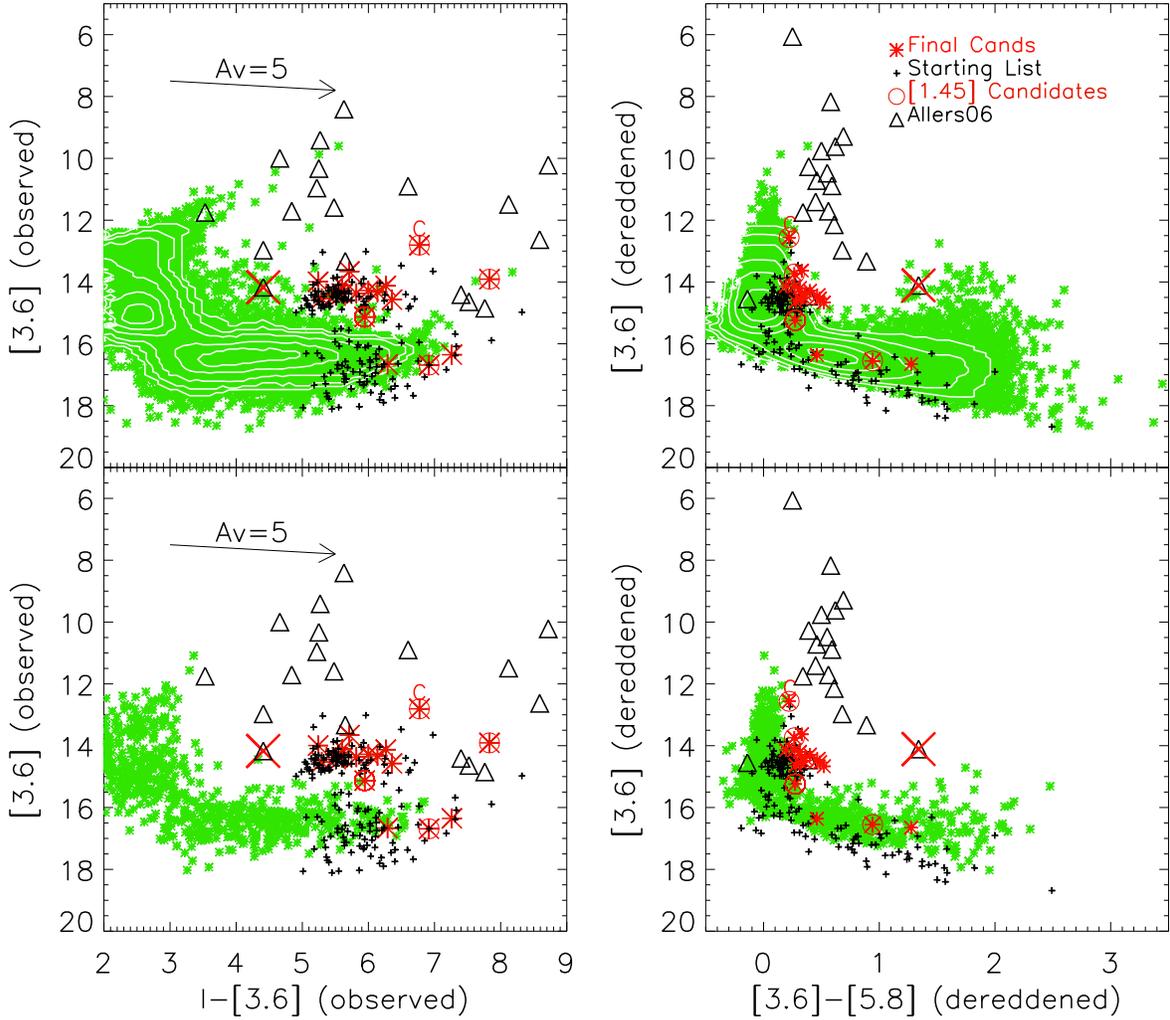}{5.5in}{0}{80}{80}{-250}{-150}
\figcaption{\label{swire}
Color magnitude diagrams showing the locations of our final candidates, as well as the original
list from which we selected these, and the corresponding colors/magnitudes from the A06
study.  The objects that fit the [1.45]-filter selection described in the text for M dwarfs are surrounded
by red circles.  Objects from the Elais N1 field of the SWIRE Spitzer Legacy program are shown in green; the upper panels
show the full sample, while the lower ones show a sub-sample with the number of randomly selected objects normalized
by the ratio of the area of our sample to that of the SWIRE Elais N1 field.  The white contours in the upper panels
are at levels of 5, 10, 20, 40, 60, and 80 percent of the peak density of extragalactic points.  The peak levels
are 24 per 0.1 mag bin in each axis in the upper left diagram and 76 per 0.1 mag bin in each axis in the upper right diagram.}
\end{figure}

Various extragalactic objects can, however, exhibit a very wide range in infrared excess.
At the faintest magnitudes within even the Spitzer {\it c2d} survey, the most serious contaminants are the
extragalactic objects, e.g. \citep{harv07}.    The IRAC magnitude limits reached in our survey are
comparable to those of the Spitzer {\it SWIRE} legacy survey \citep{lonsdale04}; the color-magnitude distribution of
these sources in the IRAC bands as shown in Figure 3 of \citet{harv07} covers a wide range, making them
very hard to eliminate with Spitzer data alone.  The SWIRE survey {\it has} acquired deep I-band photometry
of their fields, though unfortunately not matching JHK photometry.
We can examine the possible contamination of our sample by extragalactic sources by comparing our sample
colors and magnitudes with those of the SWIRE sample in bands that do overlap.  Figure \ref{swire} shows such a
comparison using the I, [3.6], and [5.8] bands.  The particular sample we have selected via the IRSA Gator
search engine\footnote{http://irsa.ipac.caltech.edu/applications/Gator/} was chosen to minimize the presence of extended sources.  In particular, we chose objects
with a ``stellarity index'' less than unity and a difference between aperture magnitude and integrated
magnitude in the I band less than 0.3.   As described in the figure, we show both the full data set
for the ElaisN1 field $\sim$ 8.5 deg$^2$, and a randomly selected subsample diluted by the ratio of the area
of our survey, 0.5 degree.  It is clear from the [3.6] vs I-[3.6] diagrams
that the brightest 3/4 of our sample are unlikely to be contaminated by extragalactic objects.
The fainter few objects, though, are located in regions of both color-magnitude diagrams where there is
a high concentration of extragalactic sources.  Spectroscopy will be needed to ascertain which of these is
stellar, though our preliminary work with [1.45]-filter photometry described below shows that at least some of our
candidates are likely to be cool, low-luminosity objects.

\begin{figure}
\plotfiddle{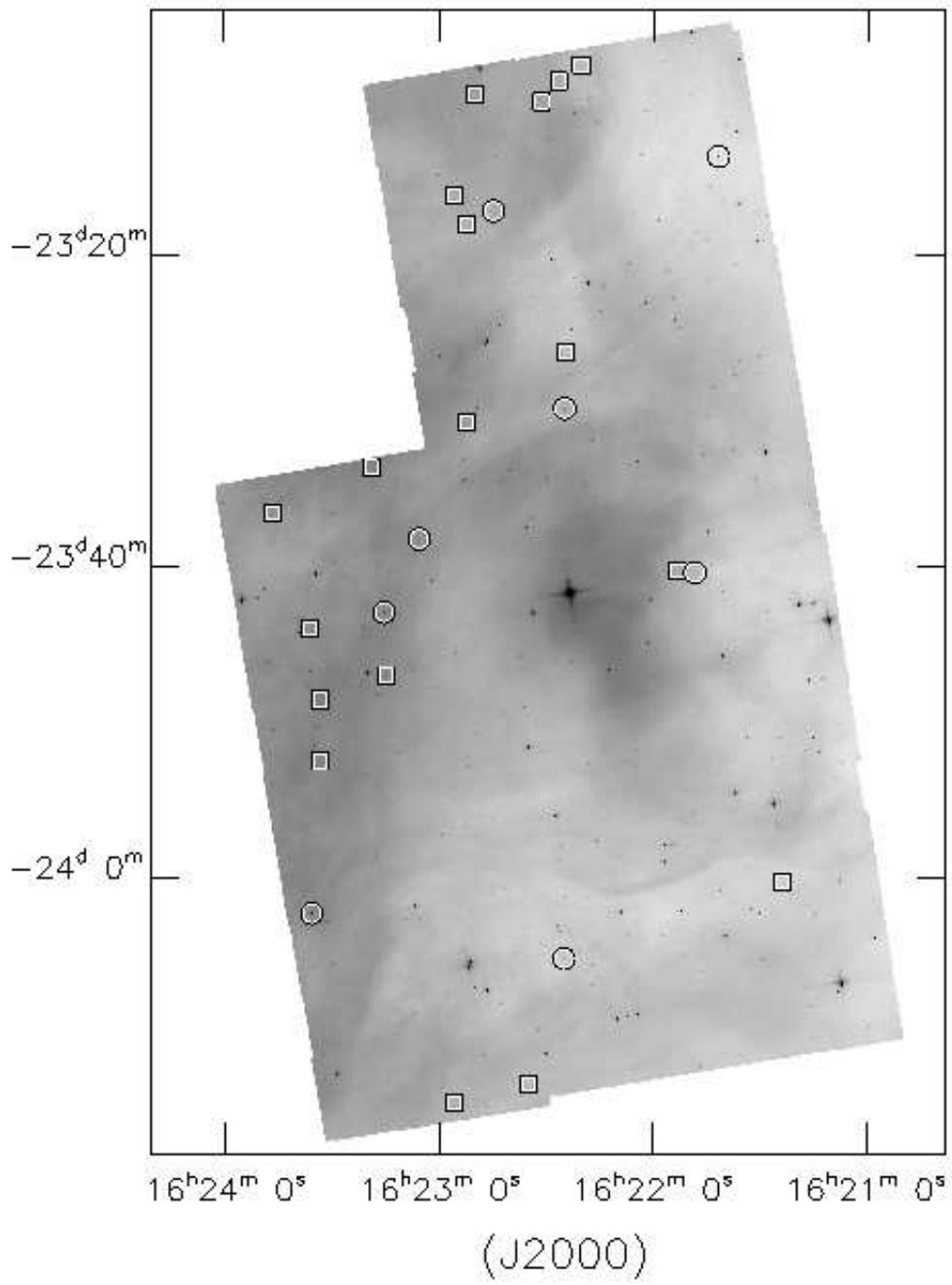}{7.0in}{0}{80}{80}{-250}{-10}
\figcaption{\label{srcdist}
Distribution of new candidate sources (boxes) and verified low-mass BD's from
A06 (circles) within our deep Spitzer search area.}
\end{figure}

A final very qualitative tool to investigate contamination is to compare the spatial distribution of
the well-vetted A06 sample of brighter low-mass BD's and M dwarfs with our new sample of candidate lower mass objects.
Figure \ref{srcdist} shows just such an image.  There is no obvious difference in the degree of clustering
or distribution within the field of the two samples.  Of course, with the relatively small number of objects
in both samples, this is not surprising.

\subsection{Verification Plans}

We have initiated several programs to determine the extent to which our new candidates do represent very
low-mass sub-stellar objects.  The first of these programs is described by \citet{allers10} who
have developed a narrow-band filter, the ``[1.45]'' filter, to identify objects with strong $H_2O$ absorption in
the H-band spectral region, indicative of very low temperature photospheres.  These data are still being
analyzed, but some preliminary results have shown that four of our candidates do indeed fit the
classification as low-photospheric-temperature objects characteristic of low-mass BD's.  These objects
have been indicated in Figures \ref{ikkf}, \ref{k58}, and \ref{swire}  by red circles.
An additional 46 objects have been identified as [1.45]-absorption candidates in our field; these were either
not detected at 5.8\micron\ or did not meet our selection criteria.

In a second program we have been awarded time on the VLT with the X-shooter spectrograph to obtain
moderate resolution spectra of the five faintest candidates in Table \ref{cand_table}.  And finally, in a collaborative
program with A. Goodman, we are hoping to obtain spectra of a number of brighter candidates with GNIRS
in the near future.  These results will also be reported in forthcoming publications.

\section{Summary}

From an initial sample of 605 objects in Ophiuchus that were selected by the original A06 photometric
criteria between I and 3.6\micron, we have identified 18 candidates that have near-infrared magnitudes
and colors consistent with very cool sub-stellar photospheres together with possible 5.8\micron\ excesses that
would be consistent with circum-object disk emission.  
Our sample extends the work of A06 to luminosities of 10$^{-4}$ \lsun, a factor of 10 below their limits.
Contamination by background field dwarfs in the brighter magnitude range of our sample and by extragalactic
objects at the fainter magnitudes is likely to be significant.  It is certainly possible that at least half
of our candidates are such contaminants.  Narrow-band filter photometry in progress, however, has shown that
at least several of our candidates are likely to be low-mass BD's with circum-object disks.
It is likely that further candidates exist in our
data set, though problems with diffuse 5.8 and 8\micron\ emission in the region make it difficult to
clearly confirm many more disk candidates.

\section{Acknowledgments}

Support for this work
was provided by NASA through RSA 1281173
 issued by the Jet Propulsion Laboratory, California Institute
of Technology,
to the University of Texas at Austin and NASA Origins grant NNX07AI83G to the University
of Texas.  P. Harvey thanks the Laboratoire Astrophysique de l'Observatoire
de Grenoble (LAOG) for its gracious support during a sabbatical while much of this research was performed.

\clearpage

\begin{table}[h]
\caption{Observations Summary (Program ID = 50025) \label{aortab}}
\vspace {3mm}
\begin{tabular}{lccc}
\tableline
\tableline
AOR  & Date & AOR Type & BCD Process \cr
\tableline


\dataset{ads/sa.spitzer\#0025243648} & 2009-04-24 & IracMap & S18.7.0 \\
\dataset{ads/sa.spitzer\#0025243904} & 2009-04-22 & IracMap & S18.7.0 \\
\dataset{ads/sa.spitzer\#0025244160} & 2009-04-23 & IracMap & S18.7.0 \\
\dataset{ads/sa.spitzer\#0025244416} & 2009-04-25 & IracMap & S18.7.0 \\
\dataset{ads/sa.spitzer\#0025244928} & 2009-04-25 & IracMap & S18.7.0 \\
\dataset{ads/sa.spitzer\#0033970176} & 2009-04-25 & IracMap & S18.7.0 \\
\dataset{ads/sa.spitzer\#0033970432} & 2009-04-23 & IracMap & S18.7.0 \\
\tableline

\tableline
\end{tabular}
\end{table}

\begin{deluxetable}{ccccccccccccc}
\tabletypesize{\footnotesize}
\rotate
\tablecolumns{11}
\tablecaption{Photometry of Selected Candidates\label{cand_table}}
\tablewidth{0pc}
\tablehead{
\colhead{Position} &
\colhead{I }  &
\colhead{J }  &
\colhead{H }  &
\colhead{K$_s$ }  &
\colhead{3.6 \micron\ }  &
\colhead{4.5 \micron\ }  &
\colhead{5.8 \micron\ }  &
\colhead{8.0 \micron\ }  &
\colhead{A$_v$ }         &
\colhead{Luminosity}  \\
\colhead{} &
\colhead{mag  }  &
\colhead{mag  }  &
\colhead{mag  }  &
\colhead{mag }  &
\colhead{mag }  &
\colhead{mag }  &
\colhead{mag }  &
\colhead{mag }  &
\colhead{mag }         &
\colhead{Log (\lsun)}  \\
}

\startdata

16 21 23.94  -24 00 19.8 &21.08$\pm$0.05 &17.53$\pm$0.03 &16.52$\pm$0.03 &15.94$\pm$0.03 &15.24$\pm$0.05 &15.11$\pm$0.05 &14.97$\pm$0.08 &14.47$\pm$0.11 &  0.0 & -3.6\\
16 21 53.36  -23 40 19.2 &23.61$\pm$0.11 &19.31$\pm$0.06 &17.94$\pm$0.04 &17.29$\pm$0.05 &16.40$\pm$0.05 &16.26$\pm$0.05 &15.92$\pm$0.19 &16.38$\pm$0.81 &  0.7 & -4.1\\
16 22 20.40  -23 07 52.3 &21.74$\pm$0.05 &16.82$\pm$0.19 &15.42$\pm$0.13 &14.89$\pm$0.12 &14.11$\pm$0.05 & \nodata &13.66$\pm$0.05 & \nodata &  6.2 & -2.8\\
16 22 24.73  -23 26 20.2 &19.98$\pm$0.05 &16.53$\pm$0.03 &15.60$\pm$0.03 &15.03$\pm$0.03 &14.58$\pm$0.05 &14.42$\pm$0.05 &14.07$\pm$0.06 &13.77$\pm$0.13 &  0.7 & -3.2\\
16 22 26.70  -23 08 53.3 &20.28$\pm$0.05 &16.79$\pm$0.19 &15.39$\pm$0.11 &14.92$\pm$0.14 &14.41$\pm$0.05 & \nodata &13.95$\pm$0.17 & \nodata &  2.1 & -3.1\\
16 22 31.60  -23 10 15.2 &19.72$\pm$0.05 &16.30$\pm$0.12 &15.35$\pm$0.10 &14.65$\pm$0.11 &14.24$\pm$0.13 & \nodata &13.92$\pm$0.07 & \nodata &  1.7 & -3.0\\
16 22 35.46  -24 13 22.2 &19.83$\pm$0.05 &16.59$\pm$0.14 &15.93$\pm$0.14 &15.13$\pm$0.18 &14.67$\pm$0.05 &14.72$\pm$0.05 &14.15$\pm$0.10 &14.03$\pm$0.14 &  0.0 & -3.3\\
16 22 50.26  -23 09 47.7 &19.89$\pm$0.05 &16.47$\pm$0.15 &15.56$\pm$0.13 &15.10$\pm$0.17 &14.48$\pm$0.05 & \nodata &14.01$\pm$0.10 & \nodata &  0.2 & -3.2\\
16 22 52.60  -23 18 07.8 &19.23$\pm$0.05 &15.79$\pm$0.03 &15.08$\pm$0.03 &14.71$\pm$0.13 &14.21$\pm$0.05 &14.04$\pm$0.05 &13.89$\pm$0.05 &13.75$\pm$0.07 &  0.1 & -3.0\\
16 22 52.77  -23 30 49.1 &20.47$\pm$0.05 &16.78$\pm$0.19 &15.72$\pm$0.14 &15.13$\pm$0.16 &14.59$\pm$0.05 &14.43$\pm$0.05 &14.20$\pm$0.06 &14.51$\pm$0.20 &  1.6 & -3.2\\
16 22 56.08  -24 14 31.2 &20.19$\pm$0.05 &16.49$\pm$0.13 &15.66$\pm$0.10 &14.96$\pm$0.15 &14.49$\pm$0.05 &14.33$\pm$0.05 &14.08$\pm$0.08 &14.29$\pm$0.15 &  1.6 & -3.1\\
16 22 56.11  -23 16 12.1 &20.93$\pm$0.06 &16.81$\pm$0.18 &15.77$\pm$0.14 &15.44$\pm$0.21 &14.73$\pm$0.05 &14.56$\pm$0.05 &14.40$\pm$0.07 &14.99$\pm$0.37 &  1.5 & -3.2\\
16 23 15.15  -23 47 05.1 &19.58$\pm$0.05 &15.12$\pm$0.03 &14.20$\pm$0.03 &13.52$\pm$0.03 &12.95$\pm$0.05 &12.73$\pm$0.05 &12.53$\pm$0.05 &12.44$\pm$0.05 &  6.4 & -2.2\\
16 23 19.28  -23 33 40.6 &19.95$\pm$0.05 &16.43$\pm$0.12 &15.69$\pm$0.12 &15.20$\pm$0.16 &14.54$\pm$0.05 & \nodata &14.22$\pm$0.06 & \nodata &  0.0 & -3.2\\
16 23 33.66  -23 52 36.5 &22.95$\pm$0.07 &19.04$\pm$0.05 &17.96$\pm$0.04 &17.24$\pm$0.04 &16.65$\pm$0.05 &16.31$\pm$0.06 &15.38$\pm$0.15 &14.13$\pm$0.21 &  0.0 & -4.1\\
16 23 33.74  -23 48 37.2 &23.60$\pm$0.10 &19.26$\pm$0.06 &18.19$\pm$0.04 &17.39$\pm$0.04 &16.56$\pm$0.06 &16.43$\pm$0.06 &15.62$\pm$0.19 & \nodata &  0.2 & -4.2\\
16 23 36.49  -23 44 04.8 &20.39$\pm$0.05 &16.60$\pm$0.16 &15.21$\pm$0.09 &15.19$\pm$0.20 &14.21$\pm$0.05 &14.07$\pm$0.05 &13.98$\pm$0.06 &14.17$\pm$0.16 &  1.2 & -3.1\\
16 23 46.86  -23 36 38.1 &19.37$\pm$0.05 &15.84$\pm$0.09 &14.92$\pm$0.07 &14.43$\pm$0.10 &13.74$\pm$0.05 & \nodata &13.35$\pm$0.05 & \nodata &  1.8 & -2.8\\

\enddata
\end{deluxetable}





\clearpage


\clearpage

\begin{thebibliography}{199}

\bibitem[Allers et al.(2006)]{allers06}
Allers, K. N., Kessler-Silacci, J. E., Cieza, L. A. \& Jaffe, D. T. 2006, \apj, 644, 364
\bibitem[Allers et al.(2007)]{allers07}
Allers, K. N. et al. 2007, \apj, 657, 511
\bibitem[Allers \& Liu(2010)]{allers10}
Allers, K.~N. \& Liu, M.~C. 2010, BAAS, 42, 335 
\bibitem[Apai et al.(2004)]{apai04}
Apai, D. et al. 2004, A\&A, 426, L53
\bibitem[Baraffe, Chabrier \& Gallardo(2009)]{baraffe09}
Baraffe, I., Chabrier, G. \& Gallardo 2009, \apj, 702, L27
\bibitem[Bate(2009)]{bate09}
Bate, M. R. 2009, MNRAS, 397, 232
\bibitem[Burrows et al.(1997)]{burrows97}
Burrows, A. et al 1997, \apj, 491, 856
\bibitem[Caballero, Burgasser \& Klement(2006)]{cabal06}
Caballero, J. A., Burgasser, A. J. \& Klement, R. 2006, A\&A, 488, 181
\bibitem[Chabrier et al.(2000)]{chabrier00}
Chabrier, G., Baraffe, I., Allard, F. \& Hauschildt, P. H. 2000, ApJ, 542, 464
\bibitem[Evans et al.(2003)]{evans03}
Evans, N. J., II, et al. 2003, PASP, 115, 965
\bibitem[Evans et al.(2007)]{evans07}
Evans, N. J., II et al. 2007, Final Delivery of Data From the c2d Legacy Project: IRAC and MIPS, http://data.spitzer.caltech.edu/popular/c2d/20071101\_enhanced\_v1/ Documents/c2d\_del\_document.pdf
\bibitem[Gully-Santiago et al.(2010)]{gully10}
Gully-Santiago, M. et al 2010, in prep.
\bibitem[Harvey et al.(2007)]{harv07}
Harvey, P. M. et al 2007, \apj, 663, 1149
\bibitem[Lonsdale et al.(2004)]{lonsdale04}
Lonsdale, C. et al. 2004, \apjs, 154, 54
\bibitem[Luhman et al.(2010)]{luhman10}
Luhman, K.L., Allen, P.R., Espaillat, C., Hartmann, L. \& Calvet, N. 2010, ApJS, 186, 111
\bibitem[Pascucci et al.(2009)]{pascucci09}
Pascucci, I. et al. 2009, \apj, 696, 143
\bibitem[Patten et al.(2006)]{patten06}
Patten, B. M. et al. 2006, \apj, 61 502
\bibitem[Scholz et al.(2009)]{scholz09}
Scholz, A., Geers, V., Jayawardhana, R., Fissel, L., Lee, E., LaFreniere, D. \& Tamura, M. 2009, \apj, 702, 805
\bibitem[Wilking, Gagn\'e \& Allen(2008)]{wilking08}
Wilking, B. A., Gagn\'e, M. \& Allen, L. E. 2008, in Handbook of Star Forming Regions, Vol. 2, ed. B. Reipurth (San Francisco: ASP), pp. 351–380






\end{thebibliography}
\end{document}